\documentclass[prl,superscriptaddress,showpacs,amssymb,amsmath,
amsfonts,aps,twocolumn,secnumarabic,floatfix]{revtex4}
\usepackage{graphics}
\usepackage{epsfig}
\begin{document} 
\bibliographystyle{try} 
 
\topmargin 0.1cm 
 
 \title{Tensor Polarization of the $\phi$ meson Photoproduced at High $t$}  
 
 
\newcommand*{\ASU }{ Arizona State University, Tempe, Arizona 85287-1504} 
\affiliation{\ASU } 

\newcommand*{\SACLAY }{ CEA-Saclay, Service de Physique Nucl\'eaire, F91191 Gif-sur-Yvette, Cedex, France} 
\affiliation{\SACLAY } 

\newcommand*{\CMU }{ Carnegie Mellon University, Pittsburgh, Pennsylvania 15213} 
\affiliation{\CMU } 

\newcommand*{\CUA }{ Catholic University of America, Washington, D.C. 20064} 
\affiliation{\CUA } 

\newcommand*{\CNU }{ Christopher Newport University, Newport News, Virginia 23606} 
\affiliation{\CNU } 

\newcommand*{\UCONN }{ University of Connecticut, Storrs, Connecticut 06269} 
\affiliation{\UCONN } 

\newcommand*{\DUKE }{ Duke University, Durham, North Carolina 27708-0305} 
\affiliation{\DUKE } 

\newcommand*{\FIU }{ Florida International University, Miami, Florida 33199} 
\affiliation{\FIU } 

\newcommand*{\FSU }{ Florida State University, Tallahassee, Florida 32306} 
\affiliation{\FSU } 

\newcommand*{\GWU }{ The George Washington University, Washington, DC 20052} 
\affiliation{\GWU } 

\newcommand*{\GBGLASGOW }{ University of Glasgow, Glasgow G12 8QQ, United Kingdom} 
\affiliation{\GBGLASGOW } 

\newcommand*{\INFNFR }{ INFN, Laboratori Nazionali di Frascati, Frascati, Italy} 
\affiliation{\INFNFR } 

\newcommand*{\INFNGE }{ INFN, Sezione di Genova, 16146 Genova, Italy} 
\affiliation{\INFNGE } 

\newcommand*{\ORSAY }{ Institut de Physique Nucleaire ORSAY, Orsay, France} 
\affiliation{\ORSAY } 

\newcommand*{\BONN }{ Institute f\"{u}r Strahlen und Kernphysik, Universit\"{a}t Bonn, Germany} 
\affiliation{\BONN } 

\newcommand*{\ITEP }{ Institute of Theoretical and Experimental Physics, Moscow, 117259, Russia} 
\affiliation{\ITEP } 

\newcommand*{\JMU }{ James Madison University, Harrisonburg, Virginia 22807} 
\affiliation{\JMU } 

\newcommand*{\KYUNGPOOK }{ Kyungpook National University, Daegu 702-701, South Korea} 
\affiliation{\KYUNGPOOK } 

\newcommand*{\MIT }{ Massachusetts Institute of Technology, Cambridge, Massachusetts  02139-4307} 
\affiliation{\MIT } 

\newcommand*{\UMASS }{ University of Massachusetts, Amherst, Massachusetts  01003} 
\affiliation{\UMASS } 

\newcommand*{\UNH }{ University of New Hampshire, Durham, New Hampshire 03824-3568} 
\affiliation{\UNH } 

\newcommand*{\NSU }{ Norfolk State University, Norfolk, Virginia 23504} 
\affiliation{\NSU } 

\newcommand*{\OHIOU }{ Ohio University, Athens, Ohio  45701} 
\affiliation{\OHIOU } 

\newcommand*{\ODU }{ Old Dominion University, Norfolk, Virginia 23529} 
\affiliation{\ODU } 

\newcommand*{\PITT }{ University of Pittsburgh, Pittsburgh, Pennsylvania 15260} 
\affiliation{\PITT } 

\newcommand*{\ROMA }{ Universita' di ROMA III, 00146 Roma, Italy} 
\affiliation{\ROMA } 

\newcommand*{\RPI }{ Rensselaer Polytechnic Institute, Troy, New York 12180-3590} 
\affiliation{\RPI } 

\newcommand*{\RICE }{ Rice University, Houston, Texas 77005-1892} 
\affiliation{\RICE } 

\newcommand*{\URICH }{ University of Richmond, Richmond, Virginia 23173} 
\affiliation{\URICH } 

\newcommand*{\SCAROLINA }{ University of South Carolina, Columbia, South Carolina 29208} 
\affiliation{\SCAROLINA } 

\newcommand*{\UTEP }{ University of Texas at El Paso, El Paso, Texas 79968} 
\affiliation{\UTEP } 

\newcommand*{\JLAB }{ Thomas Jefferson National Accelerator Facility, Newport News, Virginia 23606} 
\affiliation{\JLAB } 

\newcommand*{\UNIONC }{ Union College, Schenectady, NY 12308} 
\affiliation{\UNIONC } 

\newcommand*{\VT }{ Virginia Polytechnic Institute and State University, Blacksburg, Virginia   24061-0435} 
\affiliation{\VT } 

\newcommand*{\VIRGINIA }{ University of Virginia, Charlottesville, Virginia 22901} 
\affiliation{\VIRGINIA } 

\newcommand*{\WM }{ College of William and Mary, Williamsburg, Virginia 23187-8795} 
\affiliation{\WM } 

\newcommand*{\YEREVAN }{ Yerevan Physics Institute, 375036 Yerevan, Armenia} 
\affiliation{\YEREVAN } 


\newcommand*{\NOWNCATU }{ North Carolina Agricultural and Technical State University, Greensboro, NC 27411}

\newcommand*{\NOWGBGLASGOW }{ University of Glasgow, Glasgow G12 8QQ, United Kingdom}

\newcommand*{\NOWJLAB }{ Thomas Jefferson National Accelerator Facility, Newport News, Virginia 23606}

\newcommand*{\NOWSCAROLINA }{ University of South Carolina, Columbia, South Carolina 29208}

\newcommand*{\NOWFIU }{ Florida International University, Miami, Florida 33199}

\newcommand*{\NOWOHIOU }{ Ohio University, Athens, Ohio  45701}

\newcommand*{\NOWCMU }{ Carnegie Mellon University, Pittsburgh, Pennsylvania 15213}

\newcommand*{\NOWINDSTRA }{ Systems Planning and Analysis, Alexandria, Virginia 22311}

\newcommand*{\NOWASU }{ Arizona State University, Tempe, Arizona 85287-1504}

\newcommand*{\NOWCISCO }{ JBISoft Inc., 3877 Fairfax Ridge Rd., Suite 450, Fairfax, Virginia 22030}

\newcommand*{\NOWdeceased }{ Deceased}

\newcommand*{\NOWUK }{ University of Kentucky, LEXINGTON, KENTUCKY 40506}

\newcommand*{\NOWINFNFR }{ INFN, Laboratori Nazionali di Frascati, Frascati, Italy}

\newcommand*{\NOWUNCW }{ North Carolina}

\newcommand*{\NOWHAMPTON }{ Hampton University, Hampton, VA 23668}

\newcommand*{\NOWTulane }{ Tulane University, New Orleans, Lousiana  70118}

\newcommand*{\NOWKYUNGPOOK }{ Kyungpook National University, Daegu 702-701, South Korea}

\newcommand*{\NOWCUA }{ Catholic University of America, Washington, D.C. 20064}

\newcommand*{\NOWGEORGETOWN }{ Georgetown University, Washington, DC 20057}

\newcommand*{\NOWJMU }{ James Madison University, Harrisonburg, Virginia 22807}

\newcommand*{\NOWCALTECH }{ California Institute of Technology, Pasadena, California 91125}

\newcommand*{\NOWMOSCOW }{ Moscow State University, General Nuclear Physics Institute, 119899 Moscow, Russia}

\newcommand*{\NOWVIRGINIA }{ University of Virginia, Charlottesville, Virginia 22901}

\newcommand*{\NOWYEREVAN }{ Yerevan Physics Institute, 375036 Yerevan, Armenia}

\newcommand*{\NOWUCLA }{ University of California at Los Angeles, Los Angeles, California  90095-1547}

\newcommand*{\NOWRICE }{ Rice University, Houston, Texas 77005-1892}

\newcommand*{\NOWBATES }{ MIT-Bates Linear Accelerator Center, Middleton, MA 01949}

\newcommand*{\NOWODU }{ Old Dominion University, Norfolk, Virginia 23529}

\newcommand*{\NOWVSU }{ Virginia State University, Petersburg,Virginia 23806}

\newcommand*{\NOWORST }{ Oregon State University, Corvallis, Oregon 97331-6507}

\newcommand*{\NOWMIT }{ Massachusetts Institute of Technology, Cambridge, Massachusetts  02139-4307}

\newcommand*{\NOWCNU }{ Christopher Newport University, Newport News, Virginia 23606}

\newcommand*{\NOWGWU }{ The George Washington University, Washington, DC 20052}

\newcommand*{\NOWINFNGE }{ INFN, Sezione di Genova, 16146 Genova, Italy}

\newcommand*{\NOWNAVAL }{ Naval Research Laboratory, Washington, DC}

\newcommand*{\NOWRUTGERS }{ Rutgers, The State University of New Jersey, Piscataway, New Jersey
08854}

  
\author{K.~McCormick}
      \altaffiliation[Current address:]{\NOWRUTGERS}
     \affiliation{\SACLAY}
\author{G.~Audit}
     \affiliation{\SACLAY}
\author{J.M.~Laget}
     \affiliation{\SACLAY}
\author{G.~Adams}
     \affiliation{\RPI}
\author{P.~Ambrozewicz}
     \affiliation{\FIU}
\author{E.~Anciant}
     \affiliation{\SACLAY}
\author{M.~Anghinolfi}
     \affiliation{\INFNGE}
\author{B.~Asavapibhop}
     \affiliation{\UMASS}
\author{T.~Auger}
     \affiliation{\SACLAY}
\author{H.~Avakian}
     \affiliation{\JLAB}
     \affiliation{\INFNFR}
% ...

\collaboration{The CLAS Collaboration}
     \noaffiliation
 
\date{\today} 
 
\begin{abstract} 
As part of a measurement~\cite{phiprl} of the cross section
of $\phi$ meson photoproduction to high momentum transfer, we measured the 
polar angular decay distribution of the outgoing $K^+$ in the 
channel $\phi \rightarrow K^+K^-$ in the $\phi$ center-of-mass 
frame (the helicity frame).  We find that $s$-channel helicity 
conservation (SCHC) holds in the kinematical range where $t$-channel
exchange dominates (up to $-t \sim 2.5$~GeV$^2$ for $E_{\gamma}$~=~3.6 GeV).
Above this momentum, $u$-channel production of a $\phi$ 
meson dominates and induces a violation of SCHC.
The deduced value of the $\phi NN$ coupling constant lies in the upper range of
previously reported values.
\end{abstract} 
 
 
\maketitle 
  

The photoproduction of $\phi$ mesons at high momentum transfer $-t=-(k_{\gamma}-
k_{\phi})^2$ preferentially selects two-gluon exchange mechanisms~\cite{phiprl,mendez}. The reason is that valence
quark exchange mechanisms are strongly suppressed due to the dominant $s
\bar{s}$ component of the $\phi$ meson and the conjectured small strange
component in the nucleon wave function. However, a small non--strange component
in the $\phi$ meson wave function (as revealed by the $\phi \rightarrow \pi
\gamma$ radiative decay) or a possible strange component in the nucleon wave
function make possible a direct coupling of the $\phi$ meson to the nucleon,
allowing nucleon exchange mechanisms to show up in a well defined part of the
phase space. This happens at backward angles (high $-t$ but low $-u= -(2 m_N^2 +
m_{\phi}^2 -s -t)$), as demonstrated by  the differential cross section that 
was previously reported in Ref.~\cite{phiprl}. 

In this paper we report on the analysis of the angular decay distribution of the 
$\phi$ meson over the entire range of momentum transfer accessible at 
$E_{\gamma}$= 3.6 GeV. This was made possible by  the combined use of the 
large acceptance of the CLAS (CEBAF Large Acceptance Spectrometer) detector~\cite{CLAS} and of the
intense tagged-photon beam~\cite{tagger} at Jefferson
Laboratory (JLab). For the first time momentum transfers as large as $-t= 4$~GeV$^2$ have been reached, in contrast with a
previous measurement~\cite{dares} that was restricted to values below 1~GeV$^2$. 
This decay angular distribution adds more constraints to the reaction mechanism,
since two gluon exchange is expected to satisfy 
$s$-Channel Helicity Conservation (SCHC), while $u$-channel nucleon exchange is not. It leads 
to the determination of the $\phi NN$ coupling constant, relative to  
$\omega NN$, and provides one more constraint on a possible strange
content of the nucleon.


The decay distribution of the vector meson will be presented in
its helicity system: the z-direction is chosen opposite to the 
direction of the outgoing nucleon in the vector meson rest frame.
The decay angles $\theta$ and $\phi$ are respectively 
the polar and azimuthal angles of the $K^+$.

Assuming pure $\phi$ meson production, the decay angular distribution is 
given by~\cite{Schilling},

\begin{equation}
\frac{dN}{d\cos\theta d\phi} = W^0 (\cos\theta, \phi) + \sum_{\alpha = 1}^3 P_{\gamma}^\alpha W^\alpha(\cos\theta,\phi).
\end{equation}

\noindent
For unpolarized photons, only the first term survives.  Integrating 
over the azimuthal angle $\phi$, the decay angular distribution 
can then be written as
\begin{equation}
\frac{dN}{d\theta} = \frac{3}{4}\sin\theta\left[\left(1-\rho_{00}^0\right)
\sin^2\theta + 2 \rho_{00}^0 \cos^2\theta \right].
\label{SCHCfit_1}
\end{equation}

The tensor polarization matrix element $\rho_{00}^0$ describes 
the probability that a longitudinally polarized $\phi$ meson is produced by a transverse real 
photon.  If SCHC holds then this term is zero and there is no contribution from
longitudinally polarized $\phi$'s; in this case the angular distribution exhibits a
$\sin^3\theta$ dependence.

The $\phi$ meson may interfere with the underlying $K^+K^-$ continuum. 
Assuming an S-wave continuum (as expected to dominate near the threshold), the decay 
angular distribution becomes

\begin{eqnarray}
\label{SCHCfit}
\frac{dN}{d\theta} &=& \frac{3}{4}\left[\left(1-\rho_{00}^0\right)\sin^3\theta + 2 \rho_{00}^0 \cos^2\theta \sin\theta\right] \\ \nonumber
&+&\alpha\cos\theta\sin\theta+\frac{1}{2}\kappa\sin\theta
\end{eqnarray}
The second term, proportional to $\alpha$, describes the interference between any longitudinal 
$\phi$'s and the S-wave $K^+K^-$ continuum while the third describes the
S-wave continuum of strength $\kappa$.

As described in the following sections, the angular decay distribution
of the $K^+$ in the $\phi$ meson center-of-mass frame was measured for eight
bins in {\it t}.  The resulting distributions were fitted with Eq.~\ref{SCHCfit}
to extract values of $\rho_{00}^0$ and $\alpha$.
  

For this measurement, a 4.1 GeV electron 
beam was incident on a gold radiator of 10$^{-4}$ radiation lengths, producing
a bremsstrahlung photon beam that was tagged in the 
energy range of 3.3 -- 3.9 GeV.  A photon tagging 
system~\cite{tagger} detected the scattered electrons, with a 
resolution of 0.1\% of the incident beam energy.  The photon beam was
then incident on a liquid-hydrogen target, contained in a mylar cylinder 6 cm in
diameter and 18 cm long, which was maintained at 20.4~$^{\circ}$K.
The photon flux was determined with a pair spectrometer located
downstream of the target, which was calibrated via comparison to a 
total absorption counter.  
	
	The hadrons were detected in the CLAS spectrometer~\cite{CLAS}.  
The toroidal field of CLAS is generated with a six-coil superconducting 
magnet, effectively leading to six independent
spectrometers capable of measuring particles with polar angles between
10$^{\circ}$ and 140$^{\circ}$, thus detecting a large fraction of the 
$\phi$ mesons produced at high $-t$.  Particle momenta are determined via
magnetic analysis using trajectories reconstructed in the drift chambers~\cite{DRIFT},
and particle identification is accomplished via time-of-flight techniques using
scintillators~\cite{SCINT} that surround the toroid.  

The $\phi$ meson decay into a $K^+K^-$ pair was identified through the missing
mass in the reaction $\gamma p \rightarrow pK^+(X)$.  This technique
is preferable to measuring the $K^-$ directly, since, due to the 
magnetic field configuration, negative particles can be deflected into
the inert forward region of CLAS where they are lost.

\begin{figure}[h]
\centerline{\epsfxsize=3.5in\epsfbox{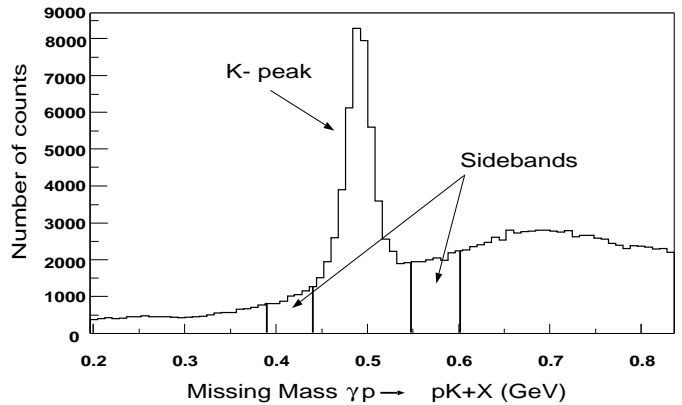}}   
\caption{Missing mass in the reaction $\gamma p \rightarrow pK^+(X)$.
\label{kminus}}
\end{figure}

Figure~\ref{kminus} shows the missing mass in the reaction
$\gamma p \rightarrow pK^+(X)$.  A well-defined $K^-$ peak can be 
seen above a background that corresponds to a combination of misidentified
pions, the contributions of multiparticle channels and accidentals
between CLAS and the tagger.  The multiparticle background 
is thought to come mostly from multipion channels such 
as $\gamma p \rightarrow p \pi^+ (\pi^- \pi^0)$.
For these types of channels, if the $\pi^+$ is misidentified as a $K^+$, then the
missing mass of the remaining two pion system can be close to that of a $K^-$,
leading to a background event.

\vspace*{.25in}

\begin{figure}[h]
\centerline{\epsfxsize=3.5in\epsfbox{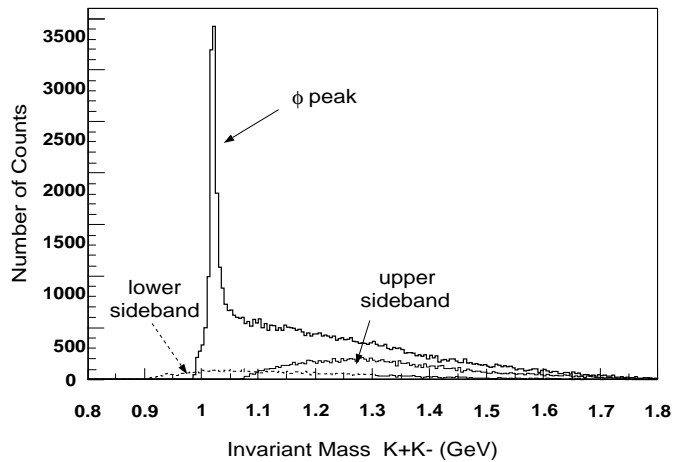}}   
\caption{Invariant Mass of $K^+(X)$.  The invariant mass of the
events that fall into the lower sideband of the $K^-$ distribution
are shown as the dotted curve, the ones of the upper sideband as 
the solid curve. 
\label{phipeak}}
\end{figure}

For each event in the $K^-$ peak, the $K^+K^-$ invariant mass is then calculated 
(see Fig.~\ref{phipeak}).  For the events whose invariant 
masses fall within the $\phi$ meson mass peak (1-1.050 GeV), the polar angle 
of the $K^+$ in the $\phi$ meson center-of-mass frame
is calculated.  As noted above, if this angle exhibits a $\sin^3\theta$ distribution, 
then SCHC holds.  The events that come from the background
under the $K^-$ peak must still be subtracted from the angular
decay distribution.  This is done by calculating the invariant
mass for each of the events in the sidebands (upper and lower,
respectively), assuming that the mass of the missing particle
is the average mass of the sideband under consideration.
The thresholds for the different reactions being considered,
$\gamma p \rightarrow p K^+ K^-$ and $\gamma p \rightarrow p K^+ X$,
vary according to the mass, as can be seen in Fig.~\ref{phipeak},
which shows the invariant mass distributions of the two sidebands.  
To take the varying kinematics into account, the contributions 
of the sidebands are taken at the same distance from their corresponding
threshold. For these events the angle of the $K^+$ in the 
helicity frame of the $K^+ X$ system is calculated, yielding an angular 
distribution for the upper and lower sidebands.  

\vspace*{.25in}

\begin{figure}[h]
\centerline{\epsfxsize=3.5in\epsfbox{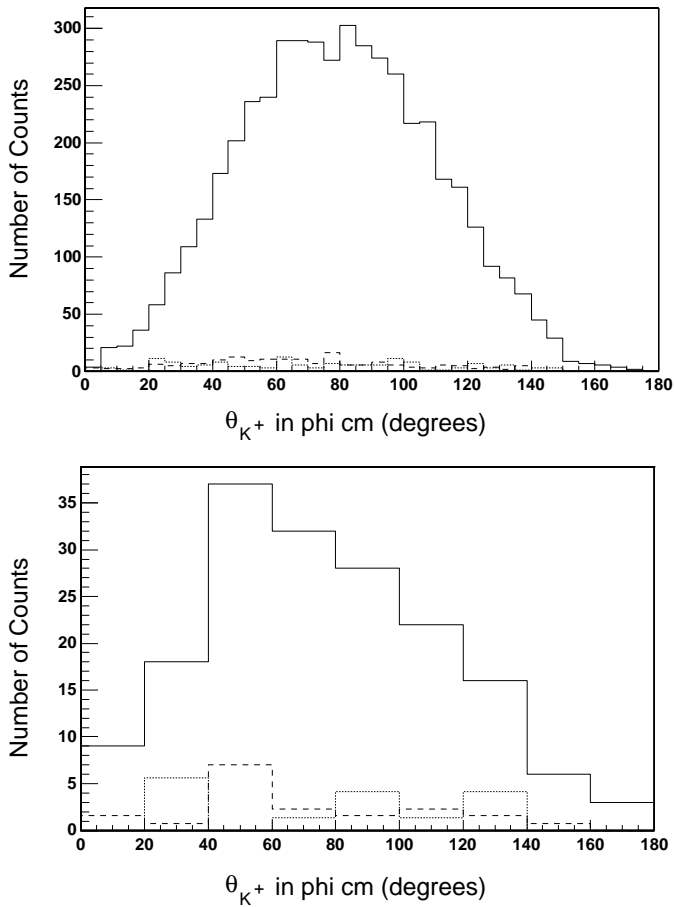}}  
\caption{The unsubtracted $\phi$ meson decay distribution (solid line) is compared
to the lower (dotted) and upper (dashed) sideband distributions. The top plot
corresponds to $\mbox\protect{0.4 < -t < 0.7}$, while the bottom plot corresponds to
$\mbox\protect{2.7<-t<3.5}$.
\label{sb_subs}}
\end{figure}

\begin{figure}[!th]
\centerline{\epsfxsize=3.5in\epsfbox{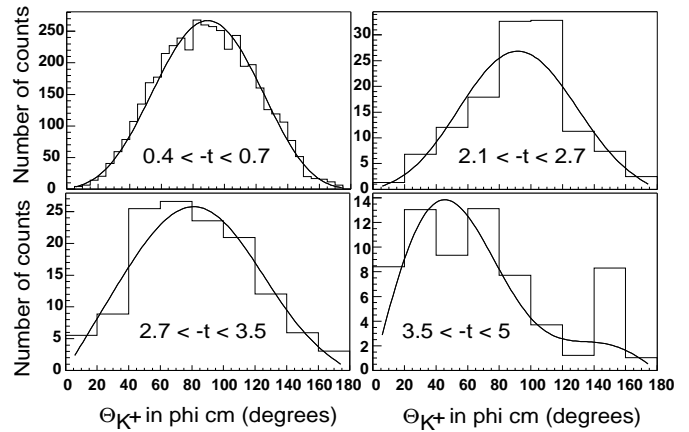}}   
\caption{The distribution of the polar angle of the $K^+$ in the $\phi$ meson center-of-mass
system for four different bins in {\it -t}, after background substraction.}
\label{phidists}
\end{figure}

Figure~\ref{sb_subs} shows the decay  angular distributions of the $\phi$ meson 
events before subtraction (solid
curves), along with the events resulting from each sideband.  
The angular distributions for the left and right sidebands are subtracted 
separately from the distribution resulting from the $\phi$ meson events. 
Within the statistics the sideband distributions are the same. At all 
values of $-t$ the background subtraction is small.  At low values of
$-t$, the sideband contribution is negligible, and at the highest values of 
$-t$ the subtraction does not significantly change the shape of the decay
distribution.  The decay distributions are then corrected for the CLAS 
efficiency, according to the method described in Ref.~\cite{phiprl}.
Some sample angular distributions are shown in Fig.~\ref{phidists} for
four different bins in {\it -t}.  All together eight bins in {\it -t} were 
measured, spanning 0.4 $\leq -t \leq$ 5.0 GeV$^2$.  


\begin{figure}[!th]
\centerline{\epsfxsize=3.in\epsfbox{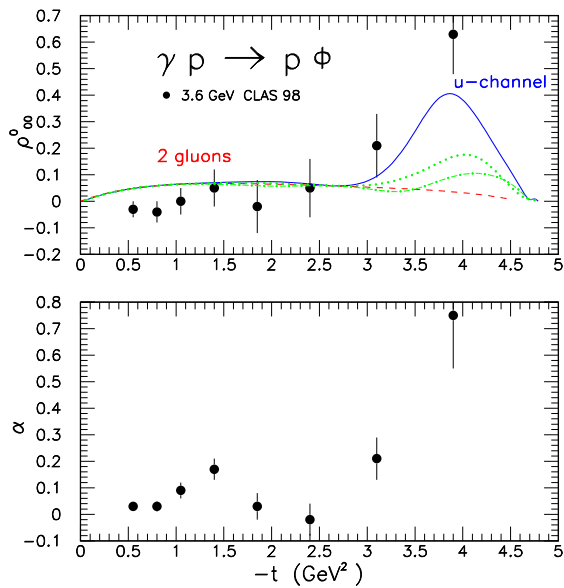}}   
\caption{The values for $\rho_{00}^0$ and $\alpha$ extracted from the decay angular
distribution of the $K^+$ in the $\phi$ meson center-of-mass system.  As described in the text,
the error bars show the spread in the values taking different hypotheses for the background
subtraction.  Also shown are the calculations of Ref.~\protect\cite{jml}.  The
dashed curve
models pomeron exchange as the exchange of two non-perturbative gluons.  The solid
curve includes the $u$-channel contribution with $g_{\phi NN}=3$. The dotted
curve corresponds to $g_{\phi NN}=1$, while the dash-dotted curve corresponds to
$g_{\phi NN}=-3$. }
\label{rho00}
\end{figure}

The extraction of $\rho_{00}^0$ and $\alpha$ was done by fitting the angular
decay distribution of the $\phi$ mesons with the left and right sideband contributions
subtracted separately for two different hypotheses for the continuum:
(see Ref.~\cite{phiprl}) either a flat contribution or a phase space distribution 
plus a contribution from the $f_0(980)$.  The ratio of the 
$\phi$ to the continuum, $\kappa$, was imposed in the 
fitting procedure and taken from the data (Fig.~5 of Ref.~\cite{phiprl}).   The four values for $\rho_{00}^0$ and $\alpha$ resulting from these
fits were then averaged.  The results are shown in Fig.~\ref{rho00}.
The error bars indicate the spread in the raw values due to the sideband and
continuum subtraction.


Also shown in Fig.~\ref{rho00} are the calculations of Ref.~\cite{jml}.  The
dashed curve
corresponds to the exchange of two non-perturbative gluons.  It dominates
the differential cross section~\cite{phiprl} and obeys SCHC.  
The distributions confirm the expectation, showing that there is essentially no 
violation of $s$-channel helicity
conservation at lower momentum transfers.  Above $-t \sim$ 2.5 GeV$^2$, the $u$-channel
contribution to $\phi$ meson production (solid line in Fig.~\ref{rho00}) begins to 
dominate and a large violation of SCHC is observed.  The values for $\alpha$ are shown in the 
lower part of the figure.  They indicate the interference between the helicity 
zero $\phi$ meson and the S-wave $K^+K^-$ continuum. It comes as no surprise that 
the interference term $\alpha$ is strongly correlated with the SCHC violation seen 
in the $\rho_{00}^0$ distribution.


The expression of the amplitude for the exchange  of the nucleon Regge trajectory
in the $u$-channel is given by Eq.~9 of~Ref.~\cite{jml}. Due to isospin considerations, 
this is the only leading trajectory that can be exchanged in the $u$-channel 
in $\phi$ as well as in $\omega$ photoproduction.  
Since the nucleon pole lies
far from the physical region, an off-shell form factor takes care of the
necessary extrapolation~\cite{thesis} and reduces the amplitude by about 33\%.
In the $\omega$ channel, all the coupling constants are known, and a fair 
agreement~\cite{jml,thesis,cano,jml1} with $\omega$ photoproduction cross sections
at backward angles is achieved with \mbox{$g_{\omega NN}= g_V(1+\kappa_V)= -15$ 
($\kappa_V$= 0)}.  This value already led to a good accounting of $\pi^0$ 
photoproduction~\cite{guidal} and falls within the range
of accepted values in the literature~\cite{vander}.

When applying this model to the $\phi$ channel, the only unknown parameter is the $\phi NN$ 
coupling constant.
The choice \mbox{$g_{\phi NN}=3$ ($\kappa_V$= 0.3)} not only leads to a good 
accounting of the rise of the differential cross section at backward 
angles~\cite{phiprl} but also of $\rho_{00}^0$ at backward angles (solid curve). 
A more robust quantity is the ratio $g_{\omega NN}/g_{\phi NN}=-5$, which gets
rid of possible uncertainties in the extrapolation to the nucleon pole.

The same value was found in the analysis of nucleon electromagnetic form 
factors~\cite{Jaffe}, as well as the analysis of nucleon-nucleon and hyperon-nucleon 
scattering~\cite{Nag}. It is higher than the value $g_{\phi NN} = 1$, 
or \mbox{$|g_{\omega NN}/g_{\phi NN}|=15$}, which is predicted  assuming 
a minimal $\omega-\phi$ mixing, as represented by the ratio of the radiative decay 
constant $\omega \rightarrow \gamma \pi$ and $\phi \rightarrow \gamma \pi$.
As shown by the dotted curve in Fig.~\ref{rho00}, this smaller value badly misses the backward peak.
Also a negative sign of the coupling constant is excluded by the data
(dash-dotted curve). This minimal $\omega-\phi$ mixing
comes from the small non-strange quark component in the $\phi$ meson wave
function (and, correspondingly,
from the small strange quark component in the $\omega$ wave function). It allows the
establishment of the relation of the coupling
constants of these mesons with the nucleon, assuming that the nucleon wave function does not contain
strange quarks. If it does, then the $\phi$ meson can couple directly to the nucleon and the coupling
constant $g_{\phi NN}$ is enhanced. The large value of the $\phi NN$ coupling constant found in this
work implies that strange quarks are present in the nucleon 
ground state wave function.



In summary, we have determined the first matrix element $\rho_{00}^0$ of the tensor 
polarization of the $\phi$, up to $-t=4$~GeV$^2$, in the full momentum transfer range available at
$E_{\gamma}=$~3.6 GeV. 
SCHC holds up to $-t= 2.5$~GeV$^2$, above which deviations point toward a value of the $\phi$ meson
nucleon coupling constant in the upper range of values already reported.

The Southeastern Universities Research Association operates the
Thomas Jefferson National Accelerator Facility under Department
of Energy contract DE-AC05-84ER40150. This work was supported in part 
by the French Commissariat \`a l'Energie Atomique, the Italian Istituto 
Nazionale di Fisica Nucleare, the U.S. Department of Energy and National 
Science Foundation, and the Korea Science and Engineering Foundation.


\begin{thebibliography}{0}
\bibitem{phiprl} E. Anciant {\it et al.}, Phys. Rev. Lett. {\bf 85}, 4682 (2000).
\bibitem{mendez} J.-M. Laget and R. Mendez-Galain, Nuc. Phys. {\bf A581}, 397
(1995).
\bibitem{CLAS} B.A. Mecking {\it et al.}, Nucl. Instr. Meth. {\bf A503}, 513 (2003).
\bibitem{tagger} D.I. Sober {\it et al.}, Nucl. Instr. Meth. {\bf A440}, 263 (2000).
\bibitem{dares} D.P. Barber {\it et al.}, Z. Phys. {\bf C12}, 1 (1982).
\bibitem{Schilling} K. Schilling, P. Seyboth and G. Wolf, Nucl. Phys. {\bf B15}, 397 (1970). 
\bibitem{DRIFT} M.D. Mestayer {\it et al.}, Nucl. Instr. Meth.  {\bf A449}, 81
(2000).
\bibitem{SCINT} E. Smith {\it et al.}, Nucl. Instr. Meth.  {\bf A432}, 265
(1999).
\bibitem{jml} J.-M. Laget, Phys. Lett. {\bf B489}, 313 (2000).
\bibitem{thesis} M. Guidal, Ph.D. Thesis, University of Orsay (1996).
\bibitem{cano} F. Cano and J.-M. Laget, Phys. Rev. {\bf D65}, 0740322 (2002).
\bibitem{jml1} J.-M. Laget, Nucl. Phys. {\bf A699}, 184c (2002).
\bibitem{guidal} M. Guidal, J.-M. Laget and M. Vanderhagen, Nucl. Phys. {\bf A627}, 644 (1997).
\bibitem{vander} M. Vanderhagen {\it et al.}, Nucl. Phys. {\bf A595}, 219 (1995).   
\bibitem{Jaffe} R.L Jaffe,  Phys. Lett.  {\bf B229}, 275  (1989).   
\bibitem{Nag} M.M Nagels, T.A Rijken and J.J de Swart, Phys. Rev. {\bf D20},  1633 (1979);
               Phys. Rev. {\bf D15},  2547 (1977); Phys. Rev. {\bf D20},  744 (1975).

\end{thebibliography}
\end{document}